\documentstyle[prl,aps,epsf,multicol]{revtex}
\begin{document}
\draft
 
\title{Measuring Fractional Charge in Carbon Nanotubes}

\author{Cristina Bena,$^1$, Smitha Vishveshwara$^1$, Leon Balents$^1$, 
  and Matthew P. A. Fisher$^2$}
\address{$^1$Department of Physics, University of California, 
Santa Barbara, CA 93106 \\
$^2$Institute for Theoretical Physics, University of California,
Santa Barbara, CA 93106--4030
}

\date{\today}
\maketitle

\begin{abstract}
  The Luttinger model of the one-dimensional Fermi gas is the
  cornerstone of modern understanding of interacting electrons in one
  dimension.  In fact, the enormous class of systems whose universal
  behavior is adiabatically connected to it are now deemed Luttinger
  liquids.  Recently, it has been shown that metallic single-walled
  carbon nanotubes are almost perfectly described by the Luttinger
  Hamiltonian.  Indeed, strongly non-Fermi liquid behavior has been
  observed in a variety of DC transport experiments, in very good
  agreement with theoretical predictions.  Here, we describe how {\sl
    fractional quasiparticle charge}, a fundamental property of
  Luttinger liquids, can be observed in impurity-induced shot noise.
\end{abstract}

\vspace{0.15cm}


\begin{multicols}{2}
\narrowtext

\emph{{\bf I. Introduction -}} The study of interacting electrons in
one dimension has opened a panoply of surprises which often seem
counter-intuitive from the perspective of higher dimensions.  The
generic behavior of one-dimensional metals is that of the {\sl
  Luttinger liquid}.\cite{Luttinger,Haldane}  In a Luttinger liquid,
interactions conspire to 
host bizarre phenomena such as the separation of spin and charge, as
well as anomalous power-law dependences in the resistivity and density
of states.  Perhaps the most fundamental difference between the
Luttinger liquid and higher-dimensional metals lies in the nature of
its quasiparticles.  The Luttinger quasiparticles are
``fractionalized'', and indeed the elementary charged quasiparticle
carries not the quantum of charge '$e$' of the electron, but instead, a
fraction '$ge$'.

While the exploration of Luttinger liquid physics began more than half
a century ago\cite{Luttinger,Haldane}, it has only found its way into the
experimental realm in the past decade. The quantum Hall system with
its chiral edge states has championed displaying Luttinger liquid
features \cite{Wen}.  In particular, it has provided the definitive
confirmation of the existence of fractional charge through shot noise
measurements\cite{FQexp}.  However, the challenge of finding a truly
one-dimensional (1D) system of interacting electrons has persisted.
A variety of transport 
experiments\cite{Dekker2,Dekker3,Bok,Dekker4,Postma}
 now convincingly demonstrate that single-walled carbon 
nanotubes(SWNTs) behave as Luttinger liquids, as predicted
theoretically\cite{KLM,Egger}.   

Here, we show that even the simple set-up of a clean armchair nanotube
with a single weak impurity is capable of flaunting a hallmark of
Luttinger liquids, namely charge fractionalization.  To understand the
physical meaning of charge fractionalization, consider a simple
thought experiment in which electrons are sequentially transferred
from a metallic electrode onto the end of a nanotube through a large
contact barrier.  With a sufficiently high barrier, the rate of charge
addition can be made very low, so that each incident electron can be
considered individually.  Immediately after the addition event, the
added charge $e$ travels as a solitonic pulse away from end of the
nanotube (incidentally, the added spin $\hbar/2$ travels as a separate
{\sl slower} soliton behind the charge).  This charge $e$ soliton may
be regarded as the remnant of the electron in the nanotube.  Upon
reaching an impurity, however, the true nature of the charge
excitations of the Luttinger liquid becomes evident.  In a
non-interacting system, an incident particle either transmits (with
probability $T$) or reflects (with probability $R=1-T$).  In the
nanotube, the charge $e$ soliton can still transmit (with probability
$T$), but the alternative possibility (which is the leading-order
scattering process occurring with probability $\approx 1-T$ for a weak
impurity) is to 'splinter' into {\sl two} solitons: a backscattered
piece of charge $ge$ and a transmitted piece of charge $(1-g)e$.  The
dimensionless ``Luttinger parameter'' $g$ ($g<1$ for repulsively
interacting Fermi systems) depends on the nature of the interactions
in the system.  In carbon nanotubes, theoretical estimates give $g
\approx 0.2$ \cite{KLM}, in good agreement with transport
measurements\cite{Dekker2,Dekker3,Bok,Dekker4,Postma}. 

Unfortunately, the difficulty of the necessary time-resolved
measurements makes the above thought experiment impractical.  Nevertheless,
as shown in the next section, the mathematics of Luttinger liquid
theory leaves no room to doubt that such strange scattering events
indeed occur.  In this paper, we determine the consequences of these
processes for {\sl shot noise}.  Strikingly, we find that at low
enough temperatures, fluctuations in the net current incident on the
weak barrier have the shot noise form appropriate for Poisson
distributed scattering events of particles of charge $ge$.
Measurements of such shot noise are more tractable, and can provide
definitive proof of charge fractionalization.

 As elucidated in what follows, to  experimentally observe
the described shot noise, we propose a four terminal 
set-up capable of measuring correlations
in current $C_I$, and the voltage drop $V$ 
{\sl across an impurity} placed along the 
nanotube. In the limit of zero temperature, we derive 
the relationship $C_I=4 g^2 {e^3 \over h} V$; $C_I$ is related to 
the backscattered current $I_B$
via $C_I=g e I_B$, while $I_B$ is given by  $I_B=4 g {e^2 \over h} V$. 
Thus one can extract the charge fraction $g$.
The SWNT, with its estimated value $g \approx 0.2$ far from unity, 
makes for an exquisite playground to test the predicted Luttinger liquid 
physics, specifically, the fractionalization of charge.

\emph{{\bf II. Formalism -}}    To extract the non-equilibrium physics of the 
set-up described above, we formulate an effective time-dependent theory 
for the bulk of the nanotube and for
the impurity site. The effective theory of the clean
single-walled nanotube in consideration may be described by the 
low energy physics of the $(N,N)$ armchair tube. In the absence 
of interactions, this involves two gapless one-dimensional metallic 
bands modeled by free fermions with linear dispersion\cite{KLM}:

\begin{equation}
H_0 = \sum_{i,\alpha}\int\! dx\ v_F[\psi^{\dagger}_{Ri\alpha}i\partial_x
\psi^{\vphantom \dagger}_{Ri\alpha}-
\psi^{\dagger}_{Li\alpha}i\partial_x\psi^{\vphantom \dagger}_{Li\alpha}],
\label{freeHam}
\end{equation}
where $v_F$ is the Fermi velocity, $R$ and $L$ label the right and left
 movers respectively, $i=1,2$ label the bands, and $\alpha = \uparrow, 
\downarrow$ the electronic spin. In this section, we set $ \hbar = e = 1$. 

The bosonized version of the fermionic operators has the form
$\psi_{R/L i \alpha} \sim e ^{i (\phi_{i \alpha} \pm \theta_{i
    \alpha})}$.  A more convenient basis, which we employ extensively,
involves a spin and channel decomposition for $\theta$: $\theta_{i,
  \rho/\sigma} = (\theta_{i \uparrow} \pm \theta_{i
  \downarrow})/\sqrt{2}$ and $\theta_{\mu \pm} = (\theta_{1 \mu} \pm
\theta_{2 \mu})/\sqrt{2}$ with $\mu = \rho,\sigma$, and a similar one
for $\phi$.  The new fields obey the canonical commutation rules $[
\phi_a(x), \theta_b(y) ] = -i \pi \delta_{a b} \Theta(x-y)$, with $a,b
= (\rho+, \rho-, \sigma+, \sigma-)$.  As discussed in Ref.\cite{KLM},
interactions effectively involve just the charge density, $\rho =
\frac{2}{\pi} \partial_x \theta_{\rho +}$.  The entire Hamiltonian
density ($H = \int\! dx\, {\cal H}$), with interactions taken into
account, then has the bosonized form:

\begin{equation}
{\cal H} = \sum_a\frac{v_a}{2 \pi}[g_a^{-1}(\partial_x\theta_a)^2 + g_a 
(\partial_x\phi_a)^2],
\label{bulk}
\end{equation}
with $g_{\rho+} \equiv g = v_F/v_{\rho+} <1$, and $g_a=1$, $v_a=v_F$
for $a \neq \rho+$.  In terms of the right- and left-moving chiral
modes $\Phi_{a}^{R/L}= g_a \phi_a \pm \theta_a$, one has associated
densities $n_{a}^{R/L}= \pm {1 \over \pi} \partial_x \Phi_a^{R/L}
$\cite{FG}. The above Hamiltonian density now takes the diagonal form:

\begin{equation}
{\cal H}=\sum_a \left\{ {{\pi v_a} \over {4 g_a}} \left[
(n_{a}^{R})^2+(n_{a}^{L})^2 \right] \right\},
\label{diagHam}
\end{equation}
with corresponding equations of motion

\begin{equation}
(\partial_t \pm v_a \partial_x)n_{a}^{R/L}=0.
\label{eqnmot}
\end{equation}

Thus, the density propagates as a one-dimensional acoustic plasmon
with renormalized velocity $v_{\rho+}$.  The parameter $g$ depends on
the ratio of the Coulomb energy between particles and the Fermi
energy, and in a SWNT has the approximate value of $0.2$\cite{KLM}.

We now consider the effect of a single weak impurity at the origin
($x=0$). In this limit, a small portion of 
quasiparticles backscatter, and the role of fractional charge is most 
transparent.  In the generic case involving no spin polarization or spin flip,
local weak backscattering processes may be described by:

\begin{eqnarray}
H_{imp} & = & \sum_{\alpha}\bigg\{ \sum_{i=1,2}u_i 
\left[ \psi^{\dagger}_{Ri\alpha}(0) \psi^{\vphantom
    \dagger}_{Li\alpha}(0)  + {\rm h.c.} \right] \nonumber \\  & &
+u_3 \left[\psi^{\dagger}_{R1\alpha}(0)
\psi^{\vphantom \dagger}_{L2\alpha}(0) + {\rm h.c.} \right] \nonumber \\  & &
+ u_4 \left[\psi^{\dagger}_{R2\alpha}(0)
\psi^{\vphantom \dagger}_{L1\alpha}(0) + {\rm h.c.}\right] \bigg\}
\label{imp}
\end{eqnarray}
where $\alpha=\uparrow, \downarrow$, 'h.c.' denotes Hermitian
 conjugation, $u_1$ and $u_2$  are weak intra-subband scattering
 potentials, and $u_3$ and $u_4$ describe the inter-subband scattering.
 Processes associated with $u_1$ and $u_2$ conserve all particle
 numbers while the $u_3$ and $u_4$ scattering terms do not conserve 
$\rho-$ and $\sigma-$ particle numbers (these arise physically for
impurities which break the sublattice-reflection symmetry of the
graphene lattice).

We note that the bosonized version of Eq.(\ref{imp}) may be expressed 
in terms of right and left moving creation and annihilation operators
 $e^{\pm i \Phi^{R/L}_a}$ 
 that describe the freely propagating chiral excitations of the system; the 
impurity site can create, destroy, or backscatter these excitations. 
Most importantly, every scattering process 
possesses a term  of the form
 $e^{\pm i \Phi^R_{\rho+}}e^{\mp i \Phi^L_{\rho+}}$, reflecting the fact that
 a  quasiparticle characterized by the creation operator
$e^{ - i \Phi^{R/L}_{ \rho+}}$ is always backscattered. 
As detailed in Ref.\cite{FG}, the operator $e^{ i \Phi^{R/L}_{ \rho+}}$ 
creates a kink of magnitude
$\pi g$ in $\Phi^R_{\rho+}$ at $x=0$, or equivalently, 
a peak in $n^{R}_{\rho+}$ of magnitude $g$. Therefore, the
magnitude of the fractional charge associated with the 
impurity backscattering is '$ge$'.

Finally we consider the real time, finite temperature action applicable at
 the impurity site.
The manner in which we employ it parallels the treatment in Ref.\cite{KF1}.
We integrate out the $\phi_a$ variables from the bulk Hamiltonian (though, 
where appropriate, we integrate out $\theta_a$ variables instead), 
and then integrate out fluctuations away from the impurity as in 
Ref.\cite{KF2}. 
Using the Keldysh approach \cite{Keldysh}, 
we write the partition function in terms of time dependent backward
 and forward paths $\theta^{\pm} \equiv \theta \pm \frac{1}{2}\tilde{\theta}$:

\begin{equation}
{\cal Z} = \int \prod_a {\cal D}\theta_a^+ {\cal D}\theta_a^- e^S,
\label{Z}
\end{equation}
with $a=(\rho+,\rho-,\sigma+,\sigma-)$. The action $S = S_0 + S_1 + S_2$ 
is given by

\begin{eqnarray}
S_0 & = & -\sum_a \left[ {\frac{1}{\pi g_a}\int d \omega \omega \coth
 \left(\frac{ \omega}{2kT}\right)|\tilde{\theta}_a(\omega)|^2 }\right. +
 \nonumber \\ 
 & & +\left. \frac{2i}{\pi g_a}\int dt \tilde{\theta_a}(t)\dot{\theta}_a(t) 
\right], \nonumber \\
S_1 & = & -i \sum_{js} \int dt [
f(\Gamma^{+}_{js}(t))-f(\Gamma^{-}_{js}(t)) ], \nonumber \\ 
S_2 & = & i \frac{2}{\pi} \int dt [ A(t) \dot{\tilde{\theta}}_{\rho+}(t) +
 \eta(t) \dot{\theta}_{\rho +}(t) ].
\label{Keldact}
\end{eqnarray}
Here,
$S_0$ describes the unperturbed system. $S_1$ is derived from the
 impurity Hamiltonian of Eq.(\ref{imp}).  The $\Gamma^\pm_{js}$
 operators are defined for $j=1\ldots 4$ and $s=\pm 1$:

\begin{eqnarray}
\Gamma_{1s}=&&\theta_{\rho+} +s\theta_{\sigma+}+\theta_{\rho-} +s
\theta_{\sigma-}, \nonumber \\
\Gamma_{2s}=&&\theta_{\rho+} +s \theta_{\sigma+}-\theta_{\rho-} -s
\theta_{\sigma-}, \nonumber \\
\Gamma_{3s}=&&\theta_{\rho+} +s\theta_{\sigma+}+\phi_{\rho-} +s
\phi_{\sigma-}, \nonumber \\
\Gamma_{4s}=&&\theta_{\rho+} +s\theta_{\sigma+}-\phi_{\rho-} -s
\phi_{\sigma-}, \label {gamma}
\end{eqnarray}
where the ${\pm}$ superscripts which denote backward and forward paths
are suppressed for all variables. $S_2$ originates from coupling the
physical current to an external source of voltage $\dot{A}$.

\emph{{\bf III. Physical Properties -}} 

\begin{figure}
\epsfxsize=2.5in
\centerline{\epsffile{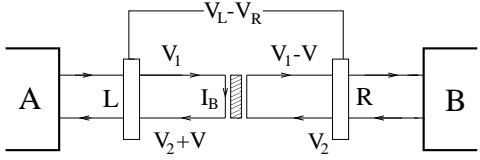}}
\vspace{0.15in}
\caption{Quasiparticle
  transport in a nanotube with a single impurity.  
The current enters the nanotubes through the 
external contacts $A$ and $B$ while the two voltage probes 
$L$ and $R$ serve to measure the voltage drop across the impurity.
Chiral modes are shown for clarity of expression, 
but they cannot be probed separately as the voltage probes 
couple to both right- and left-movers.}
\vspace{0.15in}
\label{bounce}
\end{figure}

An analysis of the nanotube with a single impurity, 
schematically shown in Fig.\ref{bounce}, serves to bring out striking
 Luttinger liquid features. However, in the absence of the impurity, 
the conductance {\sl measured across the external contacts} $A$ and
$B$, which we assume to be adiabatic, 
is $G=4 \frac{e^2}{h}$\cite{Stone}, appropriate for a non-interacting 
one-dimensional system with four channels for conductance.  
Effectively, this is due to the fact that though an isolated
nanotube would have an associated conductance $4 g {{e^2} \over {h}}$, 
the external metallic contacts are three-dimensional Fermi liquids, 
and the observed conductance involves electrons backscattering 
at the interface\cite{Stone}.

Weak backscattering in the presence of the impurity causes a reduction
in the conductance.  For a temperature $k T \gg \frac{\hbar
  v_F}{\ell}$ ($k$ is Boltzmann's constant), where $\ell$ is the distance
from the impurity to the nearest contact ($A$ or $B$), the reduction
has the form $\delta G(T) \propto -u^2 T^{2 \Delta -2}$, where $\Delta
= \frac{1}{4}(g+3)$, and `$u$' is the impurity strength\cite{FG}.
Given the estimate $g=0.2$ for the nanotube\cite{KLM}, $\delta G(T)
\propto T^{-0.4}$ ought to be observable across a wide temperature
range. Similar considerations for the limit of large tunneling
barrier, where only few electrons tunnel through, show that an
infinite wire reflects electrons completely at $T=0$.  However, at
finite temperature it exhibits the temperature dependence $G \propto
T^{2 \lambda -2}$, where $\lambda = \frac{1}{4}(\frac{1}{g}+3)$.

As seen in Fig.\ref{bounce}, the current contributing to the conductance 
involves right moving quasiparticles emerging from the left lead, and left
 movers from the right lead. As emphasized by the chiral decomposition of 
the previous section, the quasiparticles carry fractional charge $g e$, 
and a portion of them is backscattered into the lead from which they emerge. 
In the presence of an externally applied potential, the right 
and left moving chiral 
modes maintain a difference in chemical potential, $V_1-V_2-V$. 
The potential difference $V_{12}=V_1-V_2$ between the chiral modes 
arises in the presence of an external bias voltage. In the absence 
of the impurity, the current $I_0$ flowing across the wire is given by
$I_0=4 g {{e^2} \over h} V_{12}$. Unfortunately, unlike in the quantum 
Hall case, where the right- and left-movers are spatially separated, 
and $V_{12}$ is measurable\cite{FQexp},
here the leads couple to both modes. Thus, $V_{12}$ cannot be measured, 
and the ideal conductance $4 g {{e^2} \over h}$ cannot be extracted.

The voltage drop $V$ is  caused by the backscattering of quasiparticles. 
The net current traversing the wire in the presence of the impurity 
is given by
\begin{eqnarray}
I & = & I_0 - I_B \nonumber \\
& = & 4 g \frac{e^2}{h}(V_{12} - V),
\label{backsc}
\end{eqnarray}
where $V$ is the voltage drop across the impurity, and 
$I_B=4 g \frac{e^2}{h} V$ is the backscattered current. Also, as we
work in the weak backscattering limit, we have $I_B \ll I_0$.
Equation (\ref{backsc}) can be either derived from the action of 
Eq.(\ref{Keldact}) by calculating the average current
 $I = <\frac{2 e  \dot{\theta}}{\pi}>$ as the functional derivative
 $-i \frac{\delta Z}{\delta \eta}$, or by simple consideration of chiral
mode properties, as in the case of spinless
fermions\cite{FG}. In our set-up, we find that 
$V_{12}= \frac{\hbar}{e} \dot{A}$, and $V$ is the expectation value 
of the voltage operator given by

\begin{eqnarray}
\hat{V} = \frac{1}{4} \frac{h}{e} \sum_{js} u_j 
\sin(\Gamma_{js}+gA),
\label{volt}
\end{eqnarray}
with $\Gamma$'s defined in Eq.(\ref{gamma}).

The role of fractional charge is made manifest in the shot noise generated 
by the quasiparticles striking the impurity. 
To derive the general behavior of $C_I$ at finite temperature, 
we define the correlation function $C_{\cal O}$ in the quantity '${\cal O}$',
\begin{equation}
C_{\cal O}(\omega)  =  \frac{1}{2}\int \! dt\, e^{i \omega t}
<\{{\cal O}(t),{\cal O}(0) \}>,
\label{corr}
\end{equation}
and use Eq.(\ref{Keldact}) as in Ref.\cite{KF1} to obtain a relation between 
current and voltage fluctuations, $C_I$ and $C_V$ respectively. 
At low temperature, and in the limit of zero frequency, this becomes
\begin{eqnarray}
C_I = \left(\frac{4 g e^2}{h}\right)^2C_V(\omega) + 2 k T
\frac{d I}{d V_{12}} - 
\left(\frac{4 g e^2}{h}\right) 2 k T \frac{d V}{d V_{12}}.
\label{loTC}
\end{eqnarray}
We then perturbatively calculate the voltage correlations 
to lowest non-vanishing order in the impurity scattering potential to obtain
\begin{equation}
C_V(\omega \rightarrow 0) = \frac{h}{4 e}
\coth\left(\frac{g e V_{12}}{2 k T}\right)<{\hat V}>.
\label{vfl}
\end{equation}
We observe that the noise due to voltage fluctuations is partitioned
between four channels.  Putting together
Eq.(\ref{backsc}),(\ref{loTC}) and (\ref{vfl}), we obtain the desired
form of $C_I$:

\begin{eqnarray}
C_I\left(\omega \rightarrow 0\right) = && 
 g e \coth \left(\frac{g e V_{12}}{2 kT}\right) I_B + 
2 k T \frac{d I}{d V_{12}} - \nonumber \\
&&- 2 k T \frac{d I_B}{d V_{12}}. 
\label{master}
\end{eqnarray}
Setting $T=0$, we see that $C_I$ has the celebrated shot noise form $g
e I_B$. It exhibits crossover from shot noise to thermal noise when
the condition $g e V_{12} \approx 2 k T$ is satisfied.
Eq.(\ref{master}) offers a tractable starting
point for experimental data analysis.

\emph{{\bf IV. Experiment -}} 

\begin{figure}
\epsfxsize=2.5in
\centerline{\epsffile{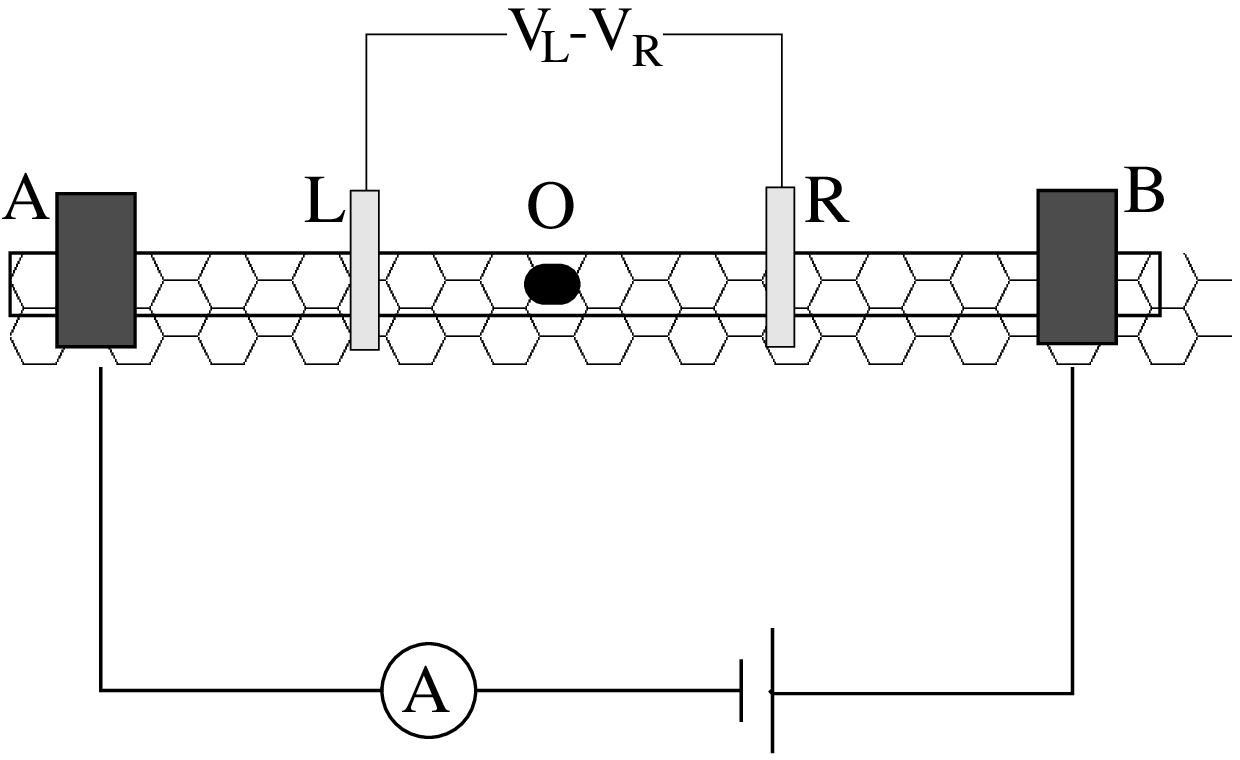}}
\vspace{0.15in}
\caption {Experimental set-up: a nanotube
  with an impurity at point 'O'  
is connected in series with an ammeter and a d.c. source
 supply. The current enters into the nanotube through
the external contacts $A$ and $B$.
A voltmeter is connected across probes $L$ and $R$.}
\vspace{0.15in}
\label{expt}
\end{figure} 

A possible experimental configuration which realizes Fig.\ref{bounce}
is shown in the four-probe geometry of Fig.\ref{expt}. As elucidated
in what follows, two sets of measurements, one in the absence of the
impurity and one in its presence, enable one to extract the shot noise
$C_I$ of Eq.(\ref{loTC}), and the voltage $V$ of Eq.(\ref{backsc})
generated across the impurity. These two quantities in turn suffice to
extract the charge fraction '$g$'.  In the proposed set-up, an
external supply maintains a voltage bias between the ends of the
nanotube and drives the current through it, while an ammeter measures
this net current $I$.  A voltmeter across $LR$ measures the difference
in potential $V_{LR}=V_L-V_R$ (see Fig.\ref{expt}).  As measurements
involve both the absence and the presence of the impurity, this
impurity would have to be created in a controlled way such as by means
of an STM tip\cite{Postma}.

The conditions in which the experiment ought to be performed are
rather specific, but feasible.  We assume that the probes $L$
and $R$ are non-invasive in their contribution to voltage drops, and
that backscattering at all leads is small.  The small backscattering
condition is equivalent to the requirement that the two-terminal
resistance of the entire wire is close to the ideal value of
$\frac{\hbar}{4 e^2} \approx 6.25 k \Omega$.  We also require the
further condition  $e V_{12}, kT \gg \frac{\hbar v_{\rho+}}{\ell}$.  
This ensures that the
one-dimensional physics of the tube is probed as opposed to that of
the three-dimensional external contacts, in essence that quasiparticle 
propagation is not coherent across the entire system.
Finally, we make the reasonable assumption that heat generated in the
external circuit is removed by phonons, hence maintaining steady
temperature $T$, and that the noise in it has the current-independent
Johnson-Nyquist form $C_{ext} \sim \frac{k T}{R_{ext}}$, where
$R_{ext}$ is the resistance of the external circuit\cite{Martinis}.

Determining $C_I$, the noise across the impurity, proves a tricky task 
due to multiple noise sources, and particularly due to possible 
correlations with scattering at the
leads. But taking into account the conditions assumed above for 
$\ell$ and for the transmission
at the leads, as well as the fact that the weak impurity also has large 
transmission, we find the noise to be additive. 
In other words, one can measure the current noise in the circuit
in the presence and absence of the impurity, $C_I^{imp}$ and $C_I^0$ 
respectively, keeping
the mean current in the circuit fixed. Then, the shot noise across the 
impurity would just be their difference; $C_I = C^{imp}_I - C^0_I$. 

One could determine the voltage $V$ generated across the impurity by
merely measuring the voltage $V_{LR}$ of Fig.\ref{expt} across it if
the leads $L$ and $R$ coupled to right- and left-movers (see
Fig.\ref{bounce}) symmetrically.  However, as shown below, even if
this coupling is asymmetric, for fixed current $I$, the voltage $V$ 
is simply the
difference between the voltages $V_{LR}$ and $V^0_{LR}$ measured
across $LR$ in the presence and in the absence of the impurity,
respectively.  To see this, let $a_R$ and $1-a_R$ be the fractions
with which the right lead couples to the right- and left- movers
respectively, and similarly for the left lead.  We {\sl assume} $a_R$
and $a_L$ are voltage-independent, a reasonable assumption given the
extremely large (${\cal O}(eV)$) electronic energy scales involved in the
microscopic matrix elements which determine this coupling.  With
reference to the voltages shown in Fig.\ref{bounce}, we then have

\begin{eqnarray}
V_R=a_R (V_1-V)+(1-a_R) V_2, \nonumber \\
V_L=a_L V_1+(1-a_L) (V_2+V), \nonumber \\ 
V_{LR}=V+(a_L-a_R)(V_1-V_2-V),
\label{vlr}
\end{eqnarray}
with the total current in the circuit given by
\begin{equation}
I=4 g \frac{e^2}{h}(V_1-V_2-V).
 \label{i}
\end{equation}
In the absence of the impurity, setting $V=0$, we have
\begin{equation}
V_{LR}^0=(a_L-a_R)(V_1^0-V_2^0),
\label{vlr0}
\end{equation}
and the corresponding current
\begin{equation}
I_0=4 g \frac{e^2}{h}(V_1^0-V_2^0).
\label{i0}
\end{equation}
Finally, for fixed current $I=I_0$,  Eqs.(\ref{vlr}-\ref{i0}) give the
required form for $V$, 
\begin{equation}
V=V_{LR}-V_{LR}^0.
\label{vimp}
\end{equation}

Now, in the limit $kT \ll V_{12}$ (which is 
of comparable magnitude to the actual applied voltage), Eq.(\ref{master}) 
takes the form $C_I = g e I_B$. Since $I_B = \frac{4 g e^2}{h} V$, we have
\begin{equation}
 C_I = 4 g^2 \frac{e^3}{h} V.
\label{result}
\end{equation}
Eq.~(\ref{result}) enables an experimental determination of '$g$',
given the measured quantities $C_I$ and $V$.  We emphasize that the
physical appearance of '$g^2$' in Eq.~(\ref{result}) has two distinct
physical origins.  As seen from the relation $I_B={{4g e^2} \over
  h}V$, one factor of $g$ simply reflects the reduced intrinsic
conductance of the Luttinger liquid.  The second factor of $g$ (in
$C_I = g e I_B$) directly follows from the fractional charge $ge$ of
the current-carrying solitons.  It is in this sense that
Eq.~(\ref{result}) is a direct measure of fractional quasiparticle charge.

In the quantum Hall system, determining fractional charge and
observing crossover from shot noise to thermal noise were performed in
a clear-cut fashion through finite temperature
measurements\cite{FQexp} that conformed to the noise described by the
analog of Eq.(\ref{master}).  Here, precise quantities cannot be
extracted due to the facts that $a_R$ and $a_L$ of Eq.'s (\ref{vlr})
and (\ref{vlr0}) are unknown, and that $V_{12}$ cannot be measured.
However, measurements at different values of the average current $I$,
and less easily at different temperatures $T$, would not only probe
quasiparticle charge, they would also display the clear
dependence of noise on thermal effects.   Also, while a gate voltage
across the Hall bar allowed one to tune the backscattering strength
elegantly, here one could change the barrier strength described in
Eq.(\ref{imp}) in a less straightforward way by studying different
impurities.

\emph{{\bf VI. Conclusion -}} The single-walled armchair nanotube
hosts a fine arena for observing many of the intriguing features of a
four channel Luttinger liquid. Its contrast with Fermi liquids becomes
strikingly apparent when its transport properties in the presence of
an impurity are compared with Landauer transport theory \cite{Land}.
Non-equilibrium measurements serve to bring out the fractional nature
of the charge carried by the quasiparticle excitations in the nanotube.

Rapid progress in experimental techniques offers scope for studying 
interacting one-dimensional systems in interesting geometries. For instance, 
a finite length tube would exhibit resonances in spectral features\cite{KLM}. 
A nanotube bearing two (or more) impurities could display resonant tunneling 
and plateaus in conductance, as observed in thin wires\cite{tunn}. Fabrication 
of nanotubes with crossed geometries\cite{cross} allows for observing quantum 
statistics of the constituent particles via the Hanbury Brown-Twiss effect as 
was studied in the quantum Hall system by Henny et. al., who used a four 
terminal geometry to measure shot noise and demonstrate the 
fluctuation-dissipation 
relation\cite{Henny}. 

Thanks to C. Dekker for inspiring this study.

This research was supported by NSF grants DMR-9985255, DMR-97-04005,
DMR95-28578, PHY94-07194 and by Broida Excellence Fellowship,
Fernando-Fithian Fellowship and Parsons Foundation Fellowship.

\end{multicols}
\end{document}